\documentclass[aps,pra,twocolumn,floatfix]{revtex4}
\usepackage{amsfonts}
\usepackage{amssymb}
\usepackage{amsmath}
\usepackage[dvips]{graphicx}
\usepackage{color}
\usepackage{amsmath}
\usepackage{graphicx}
\usepackage{amsfonts}
\usepackage{amssymb}
\usepackage{hyperref}
\usepackage{color}
\usepackage{mathrsfs}
\usepackage{isomath}
\usepackage{amsmath}
\usepackage{amsthm}
\usepackage{epstopdf}
\usepackage{txfonts}

\begin{document}

\title{Quantification of quantum steering in a Gaussian
Greenberger-Horne-Zeilinger state}
\author{Xiaowei Deng$^{1}$}
\author{Caixing Tian$^{1}$}
\author{Meihong Wang$^{1}$}
\author{Zhongzhong Qin$^{1,2}$}
\author{Xiaolong Su$^{1,2}$}
\email{suxl@sxu.edu.cn}
\affiliation{$^{1}$State Key Laboratory of Quantum Optics and Quantum Optics Devices,
Institute of Opto-Electronics, Shanxi University, Taiyuan, 030006, People's
Republic of China \\
$^{2}$Collaborative Innovation Center of Extreme Optics, Shanxi University,
Taiyuan 030006, China \\
}

\begin{abstract}
As one of the most intriguing features of quantum mechanics,
Einstein-Podolsky-Rosen (EPR) steering is a useful resource for secure
quantum networks. Greenberger-Horne-Zeilinger (GHZ) state plays important
role in quantum communication network. By reconstructing the covariance
matrix of a continuous variable tripartite GHZ state, we fully quantify the
amount of bipartite steering under Gaussian measurements. We demonstrate
that the (1+1)-mode steerability is not exist in the tripartite GHZ state,
only the collectively steerability exist between the (1+2)-mode and
(2+1)-mode partitions. These properties confirm that the tripartite GHZ
state is a perfect resource for quantum secret sharing protocol. We also
demonstrate one-way EPR steering of the GHZ state under Gaussian
measurements, and experimentally verify the introduced monogamy relations
for Gaussian steerability. Our experiment provides reference for using EPR
steering in Gaussian GHZ states as a valuable resource for multiparty
quantum information tasks.

\end{abstract}

\maketitle

\section{INTRODUCTION}

Einstein-Podolsky-Rosen (EPR) steering, the phenomenon that one party,
Alice, can steer the state of a distant party, Bob, by local measurements on
Alice is an intriguing phenomenon predicted by quantum mechanics \cite%
{Schrodinger35,Reid89,Howard07PRL,cavalcanti17review}. In the hierarchy of
quantum correlations, EPR steering represents a weaker form of quantum
nonlocality and stands between Bell nonlocality \cite{Bell65,BrunnerRMP} and
EPR entanglement \cite{entRMP}. Such correlation is intrinsically asymmetric
between the two subsystems~\cite{one-way-Theory,oneway walborn, oneway
Schneeloch, oneway Bowles, oneway opan, He15,
Adesso15,ReidJOSAB,OneWayNatPhot,OneWayPryde,OneWayGuo}, which allows
verification of EPR steering when one subsystem is untrusted~\cite{Eric09},
while the verification of entanglement is on the premise that trusted
devices are used \cite{Schrodinger35ent,entRMP}, and the Bell nonlocality
premises no need of trustiness between each other. Based on this asymmetric
feature, EPR steering is known as a potential resource for one-sided
device-independent (1sDI) quantum key distribution~\cite%
{1sDIQKD,1sDIQKD_howard,HowardOptica,CV-QKDexp,prxresource}, secure quantum
teleportation~\cite{SQT13Reid,SQT15,SQT16_LiCM}, and subchannel
discrimination \cite{subchannel}.

The one-way EPR steering with a two-mode squeezed state \cite{OneWayNatPhot}
and genuine one-way EPR steering have been experimentally demonstrated \cite%
{OneWayPryde,OneWayGuo}. It has been experimentally demonstrated that the
direction of one-way EPR steering can be actively manipulated \cite%
{ZhongzhongQin2017}, which may lead to more consideration in the
application of EPR steering. There are also some theoretical
analysis about EPR steering
among multipartite quantum state generated in different physical systems \cite%
{Yumonogamy,genuine13, Jing1, Jing2}.\ Experimental observation of
multipartite EPR steering has been reported in optical network \cite%
{ANUexp,Su2017} and photonic qubits \cite{Spainexp, USTCexp}.

The theoretical study of monogamy relations~\cite%
{ckw,Reidmonogamy,Yumonogamy,Adesso16,Shumingmonogamy} offers insights into
understanding whether and how this special type of quantum correlation can
be distributed over many different systems. In a pure three-mode Gaussian
states, the residual Gaussian steering from a monogamy inequality~\cite%
{Yumonogamy} has been demonstrated that can be used to quantify the genuine
multipartite steering~\cite{genuine13} and acquires an operational
interpretation in the context of a 1sDI quantum secret sharing protocol~\cite%
{GiannisQSS}. Very recently, the monogamy relations for EPR steering in
Gaussian cluster state has been demonstrated experimentally \cite{Su2017}.

Greenberger-Horne-Zeilinger (GHZ) state, a kind of multipartite entangled
states, is an important resource for constructing quantum network \cite%
{Loock}. For example, it has been used in quantum teleportation network \cite%
{Hide}, controlled dense coding \cite{Jing}, and quantum secret sharing \cite%
{MengGHZ,GiannisQSS}. Recently, the quantum entanglement swapping between
two Gaussian GHZ state has also been demonstrated \cite{Su2016}.

\begin{figure}[tbp]
\begin{center}
\includegraphics[width=85mm]{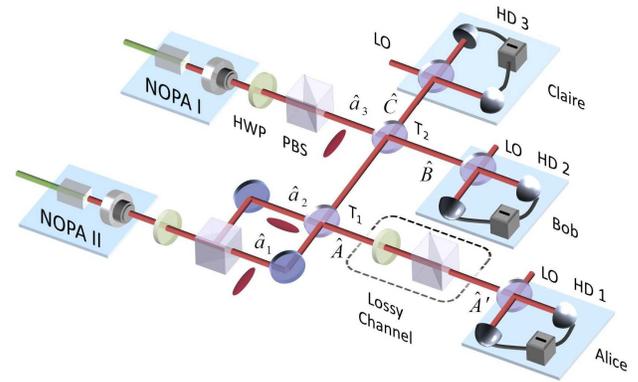}\vspace*{-0.3cm}
\end{center}
\caption{The experimental set-up. The squeezed states ($\hat{a}_{1}$, $\hat{a%
}_{2}$ and $\hat{a}_{3}$) with $-2.94$ dB squeezing at the sideband
frequency of $3$ MHz are generated from two nondegenerate optical parametric
amplifiers (NOPAs). $T_{1}$ and $T_{2}$ are the beam-splitters used to
generate the tripartite GHZ state. An optical mode ($\hat{A}$)\ of the state
is distributed over a lossy quantum channel which is composed by a half-wave
plate (HWP) and a polarization beam-splitter (PBS). HD$_{1-3}$, homodyne
detectors. LO, the local oscillator.}
\label{fig:scheme}
\end{figure}

In this paper, the distribution of quantum steering in a continuous
variable (CV) tripartite GHZ state of optical field and the monogamy
relation \cite{Reidmonogamy,Yumonogamy} of quantum steering are
theoretically and experimentally studied. With the reconstructed
covariance matrix of the GHZ state obtained from the measurement
results, the EPR steering in various bipartite splits under Gaussian
measurements are quantified. We find that a given mode of the GHZ
state cannot be steered by another mode of the state, this property
is different from those of a CV four-mode cluster state
\cite{Su2017}. However, the collective steerabilities between
(1+2)-mode and (2+1)-mode partitions are observed. We further
precisely validate the monogamy relations proposed for Gaussian
steering in the presence of
loss~\cite{Reidmonogamy,Yumonogamy,Adesso16}. This study clearly
reveals the distribution of quantum steering among different parties
in tripartite Gaussian GHZ states. Also, these characteristics
demonstrate that the CV tripartite GHZ state is a perfect resource
of quantum secret sharing protocol \cite{GiannisQSS}.

The paper is organized as follows. We present the preparation of the
tripartite GHZ state in Sec. II. The details of the experiment are presented
in Sec. III. The results and discussion are presented in Sec. IV. Finally,
we conclude the paper in Sec. V.

\section{PREPARATION OF THE TRIPARTITE GHZ STATE}

The quantum state used in the experiment is a CV tripartite GHZ entangled
state \cite{Hide,Jing,Loock}. The tripartite entangled state is prepared
deterministically by coupling a phase-squeezed state ($\hat{a}_{2}$) of
light and two amplitude-squeezed states of light ($\hat{a}_{1}$ and $\hat{a}%
_{3}$) on an optical beam-splitter network,\ which consists of two optical
beam-splitters with transmittance of $T_{1}=1/3$ and $T_{2}=1/2$,
respectively, as shown in Fig. 1. Three input squeezed states are expressed
by
\begin{align}
\hat{a}_{1}& =\frac{1}{2}\left[ e^{-r_{1}}\hat{x}_{1}^{(0)}+ie^{r_{1}}\hat{p}%
_{1}^{(0)}\right] ,  \notag \\
\hat{a}_{2}& =\frac{1}{2}\left[ e^{r_{2}}\hat{x}_{2}^{(0)}+ie^{-r_{2}}\hat{p}%
_{2}^{(0)}\right] ,  \notag \\
\hat{a}_{3}& =\frac{1}{2}\left[ e^{-r_{3}}\hat{x}_{3}^{(0)}+ie^{r_{3}}\hat{p}%
_{3}^{\left( 0\right) }\right] ,
\end{align}%
where $r_{i}$ ($i=1,2,3$) is the squeezing parameter, $\hat{x}=\hat{a}+\hat{a%
}^{\dag }$ and $\hat{p}=(\hat{a}-\hat{a}^{\dag })/i$ are the amplitude and
phase quadratures of an optical field $\hat{a}$, respectively, and the
superscript of the amplitude and phase quadratures represent the vacuum
state. The transformation matrix of the beam-splitter network is given by
\begin{equation}
U=\left[
\begin{array}{ccc}
\sqrt{\frac{2}{3}} & \sqrt{\frac{1}{3}} & 0 \\
-\sqrt{\frac{1}{6}} & \sqrt{\frac{1}{3}} & \sqrt{\frac{1}{2}} \\
-\sqrt{\frac{1}{6}} & \sqrt{\frac{1}{3}} & -\sqrt{\frac{1}{2}}%
\end{array}%
\right] ,
\end{equation}%
the unitary matrix can be decomposed into a beam-splitter network $%
U=B_{23}(T_{2})I_{2}(-1)B_{12}(T_{1}),$ where $B_{kl}(T_{j})$ stands for the
linearly optical transformation on $j-$th beam-splitter with transmission of
$T_{j}$ ($j=1,2$), where $\left( B_{kl}\right) _{kk}=\sqrt{1-T},\left(
B_{kl}\right) _{kl}=\left( B_{kl}\right) _{lk}=\sqrt{T},\left( B_{kl}\right)
_{ll}=-\sqrt{1-T},$ are matrix elements of the beam-splitter. $%
I_{k}(-1)=e^{i\pi }$ corresponds to a $180%
{{}^\circ}%
$ rotation in phase space. The output modes from the optical beam-splitter
network are
\begin{align}
\hat{A}& =\sqrt{\frac{2}{3}}\hat{a}_{1}+\sqrt{\frac{1}{3}}\hat{a}_{2},
\notag \\
\hat{B}& =-\sqrt{\frac{1}{6}}\hat{a}_{1}+\sqrt{\frac{1}{3}}\hat{a}_{2}+\sqrt{%
\frac{1}{2}}\hat{a}_{3},  \notag \\
\hat{C}& =-\sqrt{\frac{1}{6}}\hat{a}_{1}+\sqrt{\frac{1}{3}}\hat{a}_{2}-\sqrt{%
\frac{1}{2}}\hat{a}_{3},
\end{align}%
respectively. Here, we have assumed that three squeezed states have the
identical squeezing parameter ($r_{1}=r_{2}=r_{3}=r$). In experiments, the
requirement is achieved by adjusting the two nondegenerate optical
parametric amplifiers (NOPAs) to operate at same conditions. For our
experimental system, we have $r=0.339$, which corresponding to $-2.94$ dB
squeezing. The correlation variances between the amplitude and phase
quadratures of the tripartite entangled state are expressed\emph{\ }by $%
\bigtriangleup ^{2}\left( \hat{x}_{A}-\hat{x}_{B}\right) =\bigtriangleup
^{2}\left( \hat{x}_{A}-\hat{x}_{C}\right) =\bigtriangleup ^{2}\left( \hat{x}%
_{B}-\hat{x}_{C}\right) =2e^{-2r}$ and $\bigtriangleup ^{2}\left( \hat{p}%
_{A}+\hat{p}_{B}+\hat{p}_{C}\right) =3e^{-2r}$, respectively.

\begin{figure*}[tbp]
\begin{center}
\includegraphics[width=150mm]{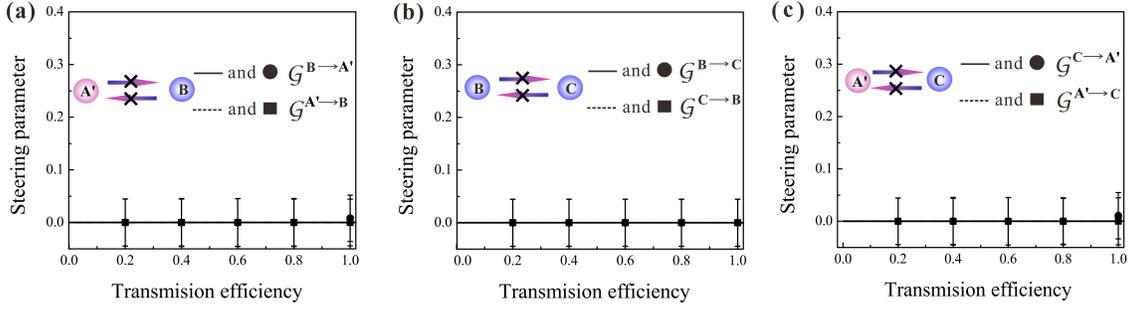}
\end{center}
\caption{EPR steering between any two modes of the GHZ state. (a - c) There
are no EPR steering between modes $\hat{A}^{\prime }$ and $\hat{B}$, $\hat{B}
$ and $\hat{C}$, $\hat{A}^{\prime }$ and $\hat{C}$, respectively, under
Gaussian measurements. In all the panels, the quantities plotted are
dimensionless. The dots and squares represent the experimental data measured
at different transmission efficiencies. Error bars represent $\pm $ one
standard deviation and are obtained based on the statistics of the measured
covariance matrices.}
\label{fig:2modes}
\end{figure*}

In the experiment, the optical mode $\hat{A}$ of the state is distributed in
a lossy channel, which is mimicked by a beam-splitter composed by a
half-wave plate and a polarization beam-splitter (Fig.~1). The output mode
is given by $\hat{A}^{\prime }=\sqrt{\eta }\hat{A}+\sqrt{1-\eta }\hat{%
\upsilon}$, where $\eta $ and $\hat{\upsilon}$\ represent the transmission
efficiency of the quantum channel and the vacuum mode induced by loss into
the quantum channel, respectively. The bipartite steerabilities among modes $%
\hat{A}^{\prime }$, $\hat{B}$ and $\hat{C}$ are investigated.

The properties of a ($n_{A}+m_{B}$)-mode Gaussian state $\rho _{AB}$ of a
bipartite system can be determined by its covariance matrix
\begin{equation}
\sigma _{AB}=\left(
\begin{array}{cc}
A & C \\
C^{\top } & B%
\end{array}%
\right) ,  \label{eq:CM}
\end{equation}%
with elements $\sigma _{ij}=\langle \hat{\xi}_{i}\hat{\xi}_{j}+\hat{\xi}_{j}%
\hat{\xi}_{i}\rangle /2-\langle \hat{\xi}_{i}\rangle \langle \hat{\xi}%
_{j}\rangle $, where $\hat{\xi}\equiv (\hat{x}_{1}^{A},\hat{p}_{1}^{A},...,%
\hat{x}_{n}^{A},\hat{p}_{n}^{A},\hat{x}_{1}^{B},\hat{p}_{1}^{B},...,\hat{x}%
_{m}^{B},\hat{p}_{m}^{B})$ is the vector of the amplitude and phase
quadratures of optical modes. The submatrices $A$ and $B$ are corresponding
to the reduced states of Alice's and Bob's subsystems, respectively.

Based on the covariance matrix of the state, the steerability of Bob by
Alice ($A\rightarrow B$) for a ($n_{A}+m_{B}$)-mode Gaussian state can be
quantified by~\cite{Adesso15}
\begin{equation}
\mathcal{G}^{A\rightarrow B}(\sigma _{AB})=\max \bigg\{0,\underset{j:\bar{\nu%
}_{j}^{AB\backslash A}<1}{-\sum }\ln (\bar{\nu}_{j}^{AB\backslash A})\bigg\},
\label{eqn:parameter}
\end{equation}%
where $\bar{\nu}_{j}^{AB\backslash A}$\ $(j=1,...,m_{B})$\ are the
symplectic eigenvalues of $\bar{\sigma}_{AB\backslash A}=B-C^{\mathsf{T}%
}A^{-1}C$, derived from the Schur complement of $A$\ in\ the covariance
matrix $\sigma _{AB}$. The quantity $G^{A\rightarrow B}$\ is a monotone
under Gaussian local operations and classical communication \cite{Adesso16}
and vanishes iff the state described by $\sigma _{AB}$\ is nonsteerable by
Gaussian measurements \cite{Adesso15}. The steerability of Alice by Bob [$%
G^{B\rightarrow A}(\sigma _{AB})$] can be obtained by swapping the roles of $%
A$\ and $B$.

\section{DETAILS OF THE EXPERIMENT}

In the experiment, the $\hat{x}$-squeezed and $\hat{p}$-squeezed states are
produced by non-degenerate optical parametric amplifiers (NOPAs) pumped by a
common laser source, which is a continuous wave intracavity
frequency-doubled and frequency-stabilized Nd:YAP-LBO (Nd-doped YAlO$_{3}$
perorskite-lithium triborate) laser. The fundamental wave at 1080 nm
wavelength is used for the injected signals of NOPAs and the local
oscillators of homodyne detectors. The second-harmonic wave at 540 nm
wavelength serves as the pump field of the NOPAs, in which through an
intracavity frequency-down-conversion process a pair of signal and idler
modes with the identical frequency at 1080 nm and the orthogonal
polarizations are generated.

Each of NOPAs consists of an $\alpha $-cut type-II KTiOPO4 (KTP) crystal and
a concave mirror. The front face of KTP crystal is coated to be used for the
input coupler and the concave mirror serves as the output coupler of
squeezed states, which is mounted on a piezo-electric transducer for locking
actively the cavity length of NOPAs on resonance with the injected signal at
$1080$ nm. The transmissivities of the front face of KTP crystal at 540 nm
and 1080 nm are $21.2\%$ and $0.04\%$, respectively. The end-face of KTP is
cut to $1^{\circ }$ along y-z plane of the crystal and is antireflection
coated for both 1080 nm and 540 nm \cite{Zhou}. The transmissivities of
output coupler at 540 nm and 1080 nm are $0.5\%$ and $12.5\%$, respectively.
In our experiment, all NOPAs are operated at the parametric deamplification
situation \cite{Zhou,Su2007}. Under this condition, the coupled modes at $%
+45^{\circ }$ and $-45^{\circ }$ polarization directions are the $\hat{x}$%
-squeezed and $\hat{p}$-squeezed states, respectively \cite{Su2007}. The
quantum efficiency of the photodiodes used in the homodyne detectors are
95\%. The interference efficiency on all beam-splitters are about 99\%.

\begin{figure*}[tbp]
\begin{center}
\includegraphics[width=150mm]{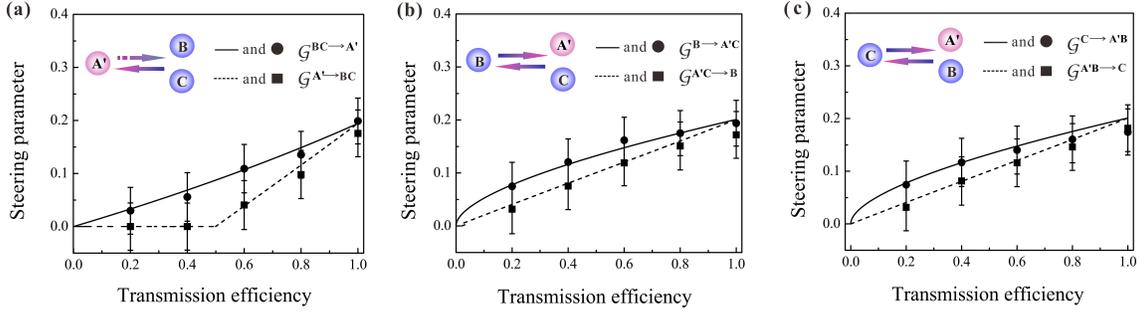}
\end{center}
\caption{EPR steering between one and two modes of the GHZ state. (b)
One-way EPR steering between modes $\hat{A}^{\prime }$ and \{$\hat{B}$, $%
\hat{C}$\} under Gaussian measurements. (b) Modes $\hat{B}$ and \{$\hat{A}%
^{\prime }$, $\hat{C}$\} can steer each other asymmetrically and the
steerability grows with increasing transmission efficiency. (c) Modes $\hat{C%
}$ and \{$\hat{A}^{\prime }$, $\hat{B}$\} can steer each other
asymmetrically and the steerability grows with increasing transmission
efficiency. In all the panels, the quantities plotted are dimensionless. The
dots and squares represent the experimental data measured at different
transmission efficiencies. Error bars represent $\pm $ one standard
deviation and are obtained based on the statistics of the measured
covariance matrices.}
\label{fig:3modes}
\end{figure*}

\begin{figure}[tbp]
\begin{center}
\includegraphics[width=85mm]{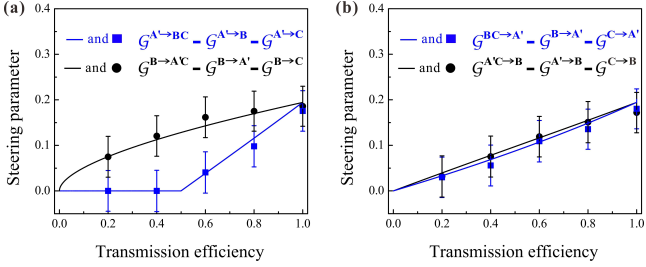}
\end{center}
\caption{Demonstration of monogamy relations. (a) Monogamy relation of
steering quantifier for $(1+2)$-mode partitions. (b) Monogamy relation of
steering quantifier for $(2+1)$-mode partitions. In both the panels, the
quantities plotted are dimensionless. The dots and squares represent the
experimental data measured at different transmission efficiencies. Error
bars represent $\pm $ one standard deviation and are obtained based on the
statistics of the measured covariance matrices.}
\end{figure}

In our experiment, the partially reconstructed covariance matrix $\sigma
_{A^{\prime }BC}$, which corresponds to the distributed mode $\hat{A}%
^{\prime }$ and modes $\hat{B}$, $\hat{C}$, is measured by three homodyne
detectors~\cite{Steinlechner}. The covariance matrix $\sigma $ is given by
\begin{equation}
\sigma =\left[
\begin{array}{ccc}
\sigma _{A^{\prime }\text{ }} & \sigma _{A^{\prime }B} & \sigma _{A^{\prime
}C} \\
\sigma _{A^{\prime }B}^{T} & \sigma _{B\text{ }} & \sigma _{BC} \\
\sigma _{A^{\prime }C}^{T} & \sigma _{BC}^{T} & \sigma _{C\text{ }}%
\end{array}%
\right] ,
\end{equation}%
Thus, the three-mode covariance matrix can be partially expressed as (the
cross correlations between different quadratures of one mode are taken as $0$%
)
\begin{align}
\sigma _{A\text{ }}& =\left[
\begin{array}{cc}
\bigtriangleup ^{2}\hat{x}_{A^{\prime }} & 0 \\
0 & \bigtriangleup ^{2}\hat{p}_{A^{\prime }}%
\end{array}%
\right] ,  \notag \\
\sigma _{B\text{ }}& =\left[
\begin{array}{cc}
\bigtriangleup ^{2}\hat{x}_{B} & 0 \\
0 & \bigtriangleup ^{2}\hat{p}_{B}%
\end{array}%
\right] ,  \notag \\
\sigma _{C\text{ }}& =\left[
\begin{array}{cc}
\bigtriangleup ^{2}\hat{x}_{C} & 0 \\
0 & \bigtriangleup ^{2}\hat{p}_{C}%
\end{array}%
\right] ,  \notag \\
\sigma _{AB\text{ }}& =\left[
\begin{array}{cc}
Cov\left( \hat{x}_{A^{\prime }},\hat{x}_{B}\right) & Cov\left( \hat{x}%
_{A^{\prime }},\hat{p}_{B}\right) \\
Cov\left( \hat{p}_{A^{\prime }},\hat{x}_{B}\right) & Cov\left( \hat{p}%
_{A^{\prime }},\hat{p}_{B}\right)%
\end{array}%
\right] ,  \notag \\
\sigma _{AC\text{ }}& =\left[
\begin{array}{cc}
Cov\left( \hat{x}_{A^{\prime }},\hat{x}_{C}\right) & Cov\left( \hat{x}%
_{A^{\prime }},\hat{p}_{C}\right) \\
Cov\left( \hat{p}_{A^{\prime }},\hat{x}_{C}\right) & Cov\left( \hat{p}%
_{A^{\prime }},\hat{p}_{C}\right)%
\end{array}%
\right] ,  \notag \\
\sigma _{BC\text{ }}& =\left[
\begin{array}{cc}
Cov\left( \hat{x}_{B},\hat{x}_{C}\right) & Cov\left( \hat{x}_{B},\hat{p}%
_{C}\right) \\
Cov\left( \hat{p}_{B},\hat{x}_{C}\right) & Cov\left( \hat{p}_{B},\hat{p}%
_{C}\right)%
\end{array}%
\right] .
\end{align}%
where $Cov\left( \hat{x},\hat{p}\right) $ is the covariance between two
corresponding quadratures. To partially reconstruct all relevant entries of
the associated covariance matrix of the state, we perform 18 different
measurements on the output optical modes. These measurements include the
amplitude and phase quadratures of the output optical modes, and the cross
correlations $\bigtriangleup ^{2}\left( \hat{x}_{A^{\prime }}-\hat{x}%
_{B}\right) $, $\bigtriangleup ^{2}\left( \hat{x}_{A^{\prime }}-\hat{x}%
_{C}\right) $, $\bigtriangleup ^{2}\left( \hat{x}_{B}-\hat{x}_{C}\right) $, $%
\bigtriangleup ^{2}\left( \hat{p}_{A^{\prime }}-\hat{p}_{B}\right) $, $%
\bigtriangleup ^{2}\left( \hat{p}_{A^{\prime }}-\hat{p}_{C}\right) $, $%
\bigtriangleup ^{2}\left( \hat{p}_{B}-\hat{p}_{C}\right) $, $\bigtriangleup
^{2}\left( \hat{x}_{A^{\prime }}+\hat{p}_{B}\right) $, $\bigtriangleup
^{2}\left( \hat{x}_{A^{\prime }}+\hat{p}_{C}\right) $, $\bigtriangleup
^{2}\left( \hat{x}_{B}+\hat{p}_{C}\right) $, $\bigtriangleup ^{2}\left( \hat{%
p}_{A^{\prime }}+\hat{x}_{B}\right) $, $\bigtriangleup ^{2}\left( \hat{p}%
_{A^{\prime }}+\hat{x}_{C}\right) $ and $\bigtriangleup ^{2}\left( \hat{p}%
_{B}+\hat{x}_{C}\right) $. The covariance elements are calculated via the
identities \cite{Steinlechner}%
\begin{align}
Cov\left( \hat{\xi}_{i},\hat{\xi}_{j}\right) & =\frac{1}{2}\left[
\bigtriangleup ^{2}\left( \hat{\xi}_{i}+\hat{\xi}_{j}\right) -\bigtriangleup
^{2}\hat{\xi}_{i}-\bigtriangleup ^{2}\hat{\xi}_{j}\right] ,  \notag \\
Cov\left( \hat{\xi}_{i},\hat{\xi}_{j}\right) & =-\frac{1}{2}\left[
\bigtriangleup ^{2}\left( \hat{\xi}_{i}-\hat{\xi}_{j}\right) -\bigtriangleup
^{2}\hat{\xi}_{i}-\bigtriangleup ^{2}\hat{\xi}_{j}\right] .
\end{align}%
In the experiment, we measured three covariance matrices for each quantum
state and got the mean steering parameters from the corresponding mean
values.

\section{RESULTS AND DISCUSSIONS}

The experimental results of the quantum steerability between any two modes [(%
$1+1$)-mode] of the tripartite GHZ state under Gaussian measurements are
shown in Figure~2. We demonstrate that there is no steering exist between
any two modes. We can account for the observation by the monogamy relation
as shown in ref. \cite{Reidmonogamy}: one mode cannot be steered by two
distinct modes simultaneously. In fact, the GHZ state is fully symmetric
under mode permutations~\cite{ckw, MengGHZ}, i.e., modes $\hat{A}$, $\hat{B}$
and $\hat{C}$ are totally symmetric. Thus, if $\hat{A}^{\prime }$ could be
steered by $\hat{B}$, it should be equally steered by $\hat{C}$, which is
forbidden by the monogamy relation \cite{Reidmonogamy}.

While, the scenario of the steering in a CV four-mode square Gaussian
cluster state shown in Ref. \cite{Su2017} shows different result. There is no
steering between any two neighboring modes of the four-mode square cluster
state, since the two neighbors of one mode are symmetric and the steering is
forbidden by the monogamy relation \cite{Reidmonogamy}. And the two-mode steering presents
between diagonal modes because in that structure any mode has just one
unique diagonal mode, thus it suffers no constraint from the monogamy
relation \cite{Reidmonogamy}.

Figure~3 shows the steerability between one mode and the other two modes of
the GHZ state, i.e., ($1+2$)-mode and ($2+1$)-mode partitions.
Interestingly, we stress that any two modes \{$\hat{\imath},\hat{\jmath}$\}
can collectively steer another mode $\hat{k}$. We further measure the
steerability when the steered party comprises two modes. Inequal
steerability between $\mathcal{G}^{BC\rightarrow A^{\prime }}$and $\mathcal{G%
}^{A^{\prime }\rightarrow BC}$ under Gaussian measurements is shown due to
the loss imposed on $\hat{A}$, and one-way steering is observed during $\eta
\in (0,0.5]$, as shown in Fig.~3(a). As shown in Fig. 3(b), modes \{$\hat{A}%
^{\prime },\hat{C}$\} and mode $\hat{B}$ can always steer each other. The
same result is observed for the steerability between modes \{$\hat{A}%
^{\prime },\hat{B}$\} and $\hat{C}$, as shown in Fig. 3(c).

Since quantum secret sharing can be implemented when the two players are
separated in a local quantum network and collaborate to decode the secret
sent by the dealer who own the other one mode \cite{GiannisQSS}. From the
results in Fig. 2 and Fig. 3, we demonstrate that steerability doesn't exist
between any two subsystems, but the collectively steerability exist between
the (1+2)-mode and (2+1)-mode partitions. These properties confirm that the
CV tripartite GHZ state is a perfect resource of quantum secret sharing
protocol.

In addition, our results $\mathcal{G}^{B\rightarrow A^{\prime }C}>0$
[Fig.~3(b)], $\mathcal{G}^{C\rightarrow A^{\prime }B}>0$ [Fig.~3(c)] and $%
\mathcal{G}^{A^{\prime }\rightarrow BC}>0$ when $\eta >0.5$ [Fig.~3(a)] also
confirm experimentally that when the steered system is composed of at least
two modes, the monogamy relation shown in ref. \cite{Yumonogamy} is
demonstrated, such that the system can be steered by more than one party
simultaneously~\cite{Yumonogamy}. With these results, we present the
experimental demonstration of the monogamy relation, Coffman-Kundu-Wootters
(CKW)-type monogamy which is seminal studied in entanglement \cite{ckw},
quantifies the distribution of the steering among different subsystems~\cite%
{Yumonogamy}. For a state contains three-mode, the CKW-type monogamy
relation presents
\begin{eqnarray}
\mathcal{G}^{k\rightarrow (i,j)}(\sigma _{ijk})-\mathcal{G}^{k\rightarrow
i}(\sigma _{ijk})-\mathcal{G}^{k\rightarrow j}(\sigma _{ijk}) &\geq &0,
\notag \\
\mathcal{G}^{(i,j)\rightarrow k}(\sigma _{ijk})-\mathcal{G}^{i\rightarrow
k}(\sigma _{ijk})-\mathcal{G}^{j\rightarrow k}(\sigma _{ijk}) &\geq &0,
\end{eqnarray}%
where $i,j,k\in \{\hat{A}^{\prime },\hat{B},\hat{C}\}$. All possible
configurational types of $(1+2)$-mode and ($2+1$)-mode steering monogamy
relation has been experimentally verified with our experimental results, as
shown in Fig.~4.

\section{CONCLUSION}

In summary, we fully quantify the steering characterization for all
bipartite configurations with the deterministically generated GHZ state and
the reconstructed covariance matrix under Gaussian measurements. The
distribution of EPR steering over (1+1)-mode, (1+2)-mode and (2+1)-mode
partitions of a CV tripartite Gaussian GHZ state subject to asymmetric loss
have been investigated. The fact that a given mode of the state cannot be
steered by another mode is observed, and the collective steerabilities
between (1+2)-mode and (2+1)-mode partitions are observed. These results
demonstrate that the CV tripartite GHZ state is a perfect resource of
quantum secret sharing protocol.

We also provide experimental confirmation for two types of monogamy
relations for the CV tripartite Gaussian GHZ state, which bound the
distribution of steerability among different modes. Our work thus provides a
concrete in-depth understanding of EPR steering in paradigmatic multipartite
states such as GHZ states, and advances our fundamental knowledge of
monogamy relations for Gaussian steerability. This work thus can be useful
in establishing secure teleportation fidelity thresholds, and bounds on 1sDI
quantum key distribution and secret sharing among many sites over lossy
quantum channels.

\textbf{Acknowledgments} This research was supported by National Natural
Science Foundation of China (Grants No. 11522433 and No. 61475092), the
program of Youth Sanjin Scholar, National Basic Research Program of China
(Grant No. 2016YFA0301402), and the Fund for Shanxi "1331 Project" Key
Subjects Construction. \newline

\end{document}